\documentclass[3p,times,twocolumn]{elsarticle}

\usepackage{ecrc}

\volume{00}

\firstpage{1}

\journalname{Nuclear Physics B Proceedings Supplement}

\jid{nuphbp}

\jnltitlelogo{Nuclear Physics B Proceedings Supplement}

\usepackage{amssymb}

\usepackage[figuresright]{rotating}
\usepackage{amsmath}

\begin{document}

\begin{frontmatter}




\title{Hadronic currents for $\tau^-\to\pi^-\pi^0\nu_\tau$ and other decays of interest in TAUOLA}


\author{Pablo Roig}

\address{Grup de F\'{\i}sica Te\`orica \& IFAE. 
Universitat Aut\`onoma de Barcelona. 
Fac. De Ciencies, Edif. Cn. 
E-08193, Bellaterra (Barcelona), Spain}

\begin{abstract}
A new set of hadronic form factors, which has been implemented in TAUOLA, is described.
\end{abstract}


\end{frontmatter}


\section{Semileptonic tau decays in TAUOLA}\label{Intro}
Tau decays including hadrons are a privileged scenario \cite{HadTau} to study non-perturbative QCD in a rather clean environment provided by the electroweak half of the process. 
In Ref.~\cite{arXiv:0912.0749} it was concluded that the most essential missing step to perform, concerning the study of hadronic tau decays, was the appropriate choice of 
the hadronic currents. The original version of the Monte Carlo (MC) generator TAUOLA \cite{CERN-TH-6793-93} used the so-called K\"uhn-Santamar\'{i}a (KS) model \cite{KS}, and its extensions, 
to construct them. Within this model, they are built to fulfil the leading order (LO) result in the low-energy effective field theory 
of QCD, Chiral Perturbation Theory ($\chi PT$) \cite{ChPT}, but they violate the next-to-leading order (NLO) one \cite{t3p}. This approach was sufficient and succesful twenty years 
ago, but already the CLEO and Aleph Collaborations realized, later on, departures of the predictions from data, a feature which could be expected taking into account several 
inconsistencies in later parametrizations of the three meson modes including Kaons \cite{arXiv:0810.1255}. This resulted in private versions of the code, with fine-tuned 
initializations -which sometimes violated basic principles of QCD\cite{hep-ph/0411333}- that were documented in Ref.~\cite{Golonka:2003xt}. Nowadays, with the massively 
increased data samples from the B-factories BaBar and Belle -the most of which have not been analyzed yet-, it is pressing to upgrade the hadronic currents in TAUOLA in order 
to obtain as much QCD information as possible from experiment; moreover with the perspective of the super-flavour factories producing huge amounts of high-quality data in the 
near future.

In Ref.~\cite{RChLTAUOLA} a new set of form factors for hadronic tau decays based in analytical results obtained from Resonance Chiral Theory ($R\chi T$)\cite{RChT} is 
documented. In these Proceedings, a description of the implementation of the new modules of the MC program for its user is given in Ref.~\cite{Was}. Here we focus on the 
hadronic currents themselves.

Lorentz invariance determines the most general decomposition of the hadronic current. This is 
\begin{equation}\label{2mesons}
 J^\mu\,=\,N \bigl[ (p_1 - p_2)^\mu F^{V}(s) + (p_1 + p_2)^\mu F^{S}(s) \bigr]\,,
\end{equation}
in the two-meson channels, where $F^{V}(s)$ and $F^{S}(s)$ are the vector and scalar form factors, given in terms of $s=(p_1+p_2)^2$, and
 \begin{eqnarray}\label{3mesons}
& & J^\mu =N \bigl\{T^\mu_\nu \bigl[ (p_2-p_3)^\nu F_1  + (p_3-p_1)^\nu F_2  \\
& & + (p_1-p_2)^\nu F_3 \bigr]+ q^\mu F_4  -{ i \over 4 \pi^2 F^2} \epsilon^\mu_{.\ \nu\rho\sigma} p_1^\nu p_2^\rho p_3^\sigma F_5 \bigr\}\,,\nonumber
\label{fiveF}
\end{eqnarray}
in the three-meson decays\footnote{The (by-far) most important four-meson tau decay, into four pions, is currently parametrized following Ref.~\cite{hep-ph/0201149}.}, 
where $T_{\mu\nu} = g_{\mu\nu} - q_\mu q_\nu/q^2$ stands for the transverse projector, and $q^\mu=(p_1+p_2+p_3)^\mu$ is the momentum of the hadronic system. Only two 
among the $F_1$, $F_2$ and $F_3$ (axial-vector) form factors are independent. $F_4$ is the generally suppressed pseudoscalar form factor and $F_5$ is the vector form 
factor. The kinematical invariants are $q^2$, and two of the three $s_{ij}=(p_i+p_j)^2=s_k,\,i\neq j\neq k=1,2,3$, since $s_3 = q^2 - s_1 - s_2 + m_1^2 + m_2^2 + m_3^2$. 
Chiral symmetry relates the $N$ factors for the diverse decay channels in Eqs.(\ref{2mesons}) and (\ref{3mesons}), respectively. Unfortunately, there is no analytic way of 
deriving the expression of these form factors from the QCD Lagrangian. This does not mean that the underlying theory is useless to find them, as we discuss next.

\section{Theoretical setting}\label{Theory}
QCD has a well-defined expansion parameter at low energies within the light quark sector, $\Lambda_{\chi PT}$, that allows to build $\chi PT$. The associated approximate 
chiral symmetry is useful to understand the data at low values of the invariant mass of the hadronic system, but it is insufficient to explain them throughout the whole 
phase space \cite{hep-ph/9604279}. This happens because $\Lambda_{\chi PT}$ is no longer small for $E\gtrsim m_K$. $1/N_C$ \cite{large-N_C} seems a good candidate to build 
the expansion upon, given its success in explaining many features of meson phenomenology \cite{hep-ph/9802419}. Furthermore, it proves efficient in dealing with effective 
field theories of QCD in the low- and intermediate-energy regions, $\chi PT$ and $R\chi T$ in our context \cite{hep-ph/0205030}. Working at lowest order in $1/N_C$ amounts 
to consider an infinite number of stable resonances which experience local effective interactions at tree level among them. This supports $R\chi T$, which includes the 
$\chi PT$ Lagrangian at LO, and reproduces the one at NLO upon integration of the resonances \cite{RChT}, which are active degrees of freedom in the theory. $R\chi T$ 
provides a rigorous way \cite{hep-ph/0003320} of computing a NLO effect in this expansion, the resonance widths, of paramount importance to understand hadronic tau decay 
data. A model independent realization of the infinite tower of resonances remains unknown. We model this setting by cutting the spectrum in a way that resembles nature. 
One hopes that since the lowest-lying resonances dominate the light-flavour Physics, this procedure would allow to capture the essentials of the involved dynamics. The 
$R\chi T$, derived from symmetries, is still lacking the short-distance QCD behavior. When it is imposed to the Green functions \cite{GFs} and associated form factors, a 
number of relations between the Lagrangian couplings arise, making the theory more predictive.

\section{$\tau^-\to\pi^-\pi^0\nu_\tau$}\label{tau2pi}
There are different approaches to deal with the diverse energy regimes which are probed through this decay: $\chi PT$ should be a valid description of the data for $s<< M_\rho$. 
Computations at NNLO are available both in the $SU(2)$ \cite{hep-ph/9805389} and in the $SU(3)$ \cite{hep-ph/0203049} symmetry cases. In the region $M_\rho\lesssim s\lesssim1$ GeV, 
the chiral expansion breaks down and the dominant $\rho(770)$ exchange has to be accounted for. Several approaches have been developed. Among them, matching $\chi PT$ results 
to vector meson dominance using an Omn\`es solution \cite{9142} for the dispersion relation \cite{arXiv:hep-ph/9707347}, employing an Omn\`es solution for the dispersion 
relation \cite{hep-ph/0101194} or utilizing the unitarization approach \cite{Unitarization}. For larger energies, in the $1$-$2$ GeV region, the excited resonances play an 
important role and shall be incorporated to the description. Ref.~\cite{hep-ph/0208199} includes the $\rho(1450)$ through a Schwinger-Dyson-like resummation and Refs.\cite{DualQCD} 
include a tower of resonances inspired from dual QCD. Since Ref.~\cite{arXiv:hep-ph/9707347} will be our starting point, let us recall their main features in the following.

In this case $N=\sqrt{2}$ in Eq.(\ref{2mesons}), and the scalar form factor is zero in the $SU(2)$ symmetry limit \footnote{Even when first order isospin violating 
corrections are included \cite{hep-ph/0104267} it does not contribute.}. The vector form factor, at NLO in $\chi PT$, is
\begin{equation}\label{Op4FF}
 F^{V}(s)=1+\frac{2L_9^r(\mu)}{F_\pi^2}s-\frac{s}{96\pi F_\pi^2}\left[A_\pi(s)+\frac{1}{2}A_K(s)\right]\,,
\end{equation}
with
\begin{eqnarray}\label{A_P function}
 & & A_P(s)=\mathrm{Log}\left(\frac{m_P^2}{\mu^2}\right)+\frac{8m_P^2}{s}-\frac{5}{3}+\sigma_P^3\mathrm{Log}\left(\frac{\sigma_P+1}{\sigma_P-1}\right)\,,\nonumber\\
 & & \sigma_P=\sqrt{1-\frac{4m_P^2}{s}}\,.
\end{eqnarray}
The computation in $R\chi T$, within the antisymmetric tensor formalism, reads
\begin{equation}
  F^{V}(s)=1+\frac{F_VG_V}{F_\pi^2}\frac{s}{M_\rho^2-s}\,.
\end{equation}
When $F^{V}(s)\xrightarrow{s\to\infty}0$ is required, the condition $F_VG_V=F_\pi^2$ is found, which yields the vector meson dominance prediction ($\mu\sim M_\rho$)
\begin{equation}\label{VMD}
  F^{V}(s)= \frac{M_\rho^2}{M_\rho^2-s}\,\Leftrightarrow L_9^r=\frac{F_\pi^2}{2M_\rho^2}\,.
\end{equation}
The matching of Eqs.(\ref{Op4FF}) and (\ref{VMD}) is straightforward
\begin{equation}\label{MatchedFF}
  F^{V}(s)= \frac{M_\rho^2}{M_\rho^2-s}-\frac{s}{96\pi F_\pi^2}\left[A_\pi(s)+\frac{1}{2}A_K(s)\right]\,.
\end{equation}
When unitarity and analiticity properties are required, the Omn\`es solution emerges
\begin{equation}\label{MatchedFF+Omnes}
 F^{V}(s)= \frac{M_\rho^2}{M_\rho^2-s}\mathrm{exp}\left\lbrace-\frac{s}{96\pi F_\pi^2}\left[A_\pi(s)+\frac{1}{2}A_K(s)\right]\right\rbrace\,.
\end{equation}
Not surprisingly, the $\rho(770)$ off-shell width is related to the imaginary part of the same loop function
\begin{equation}\label{rho width}
 \Gamma_\rho(s)=\frac{-M_\rho s}{96\pi^2F_\pi^2}\Im m\left[A_\pi(s)+\frac{1}{2}A_K(s)\right]\,.
\end{equation}
A possible solution to avoid double counting of the imaginary parts, which was adopted in Ref.~\cite{arXiv:hep-ph/9707347}, is
\begin{equation}\label{GuerreroPich}
 F^{V}(s)= \frac{M_\rho^2}{M_\rho^2-s-iM_\rho\Gamma_\rho(s)}e^{\left\lbrace-\frac{s}{96\pi F_\pi^2}\Re e\left[A_\pi(s)+\frac{1}{2}A_K(s)\right]\right\rbrace}\,.
\end{equation}
Eq.(\ref{GuerreroPich}) reproduces $\chi PT$ at NLO, vanishes at $s\to\infty$, has $SU(2)$ symmetry built-in and complies with analiticity and unitarity constraints up to 
first order in the expansion of the exponential. This description was successfully confronted to data using only one parameter, $M_\rho$. Present data have become much 
more precise and the Belle results \cite{arXiv:0805.3773} point to an interference pattern between excited resonances in this decay. All this motivates us to include \cite{Paper}, 
analogously, the contribution of the excited resonances [$\rho'=\rho(1450)$ and $\rho''=\rho(1700)$ in this case] while keeping these nice properties \cite{hep-ph/0605096}
\begin{eqnarray} \label{SU2formula}
F^V(s) = \frac{M_\rho^2+ s (\gamma e^{i \phi_1}+\delta e^{i \phi_2})}{M_\rho^2-s-i M_\rho \Gamma_\rho(s)} e^{\left\lbrace\Re e\left[ \frac{-s}{96 \pi^2 F_\pi^2} \left( A_\pi(s) + \frac{1}{2}A_K(s)\right)\right]\right\rbrace}& & \nonumber\\
-\frac{\gamma s e^{i \phi_1}}{M_{\rho'}^2-s-i M_{\rho'} \Gamma_{\rho'}(s)} e^{\left\lbrace \frac{-s \Gamma_{\rho'}\left(M_{\rho'}^2\right)}{\pi M_{\rho'}^3 \sigma_\pi^3\left(M_{\rho'}^2\right)} \Re e A_\pi(s)\right\rbrace}& & \nonumber\\
-\frac{\delta s e^{i \phi_2}}{M_{\rho''}^2-s-i M_{\rho''} \Gamma_{\rho''}(s)} e^{\left\lbrace \frac{- s \Gamma_{\rho''}\left(M_{\rho''}^2\right)}{\pi M_{\rho''}^3 \sigma_\pi^3\left(M_{\rho''}^2\right)} \Re e A_\pi(s)\right\rbrace}\,. & &
\end{eqnarray}
The parameters $\gamma$ and $\delta$ are related to $R\chi T$ couplings for the excited resonances [as $\gamma_{K\pi}$ in Eq.(\ref{VFFKpi})], the $\Gamma_{\rho'}(s)$ and 
$\Gamma_{\rho''}(s)$ widths are modeled as decays to two pions, and the phases $\phi_1$ and $\phi_2$ should vanish, at least, as $1/N_C$. Eq.(\ref{SU2formula}) corresponds to 
what is included in TAUOLA right now \footnote{The values of the parameters for this mode and the others can be found in the quoted references, where comparisons to data are 
also available.}. SU(2) breaking has only been coded partially, through the kinematical and loop functions. Electromagnetic corrections \cite{hep-ph/0104267, EMCorrs} 
have been considered \cite{Paper} but not incorporated to the MC yet. An alternative approach to using the Omn\`es solution consists in employing an n-subtracted 
dispersion relation where the relevant phaseshift, $\delta_1^1(s)$, is obtained as $\Im mF^V(s)/\Re eF^V(s)$. In this procedure \cite{arXiv:0807.4883}, unitarity and 
analiticity are satisfied to all orders with \cite{Paper}
\begin{eqnarray} \label{SU2formula_2}
F^V(s) = \frac{M_\rho^2+ s (\gamma e^{i \phi_1}+\delta e^{i \phi_2})}{M_\rho^2\left[1+\xi_\rho\Re e\left( A_\pi(s) +\frac{1}{2} A_K(s)\right)\right]-s-i M_\rho \Gamma_\rho(s)} \nonumber\\
-\frac{\gamma s e^{i \phi_1}}{M_{\rho'}^2\left[1+\xi_{\rho'} \Re e A_\pi(s)\right]-s-i M_{\rho'} \Gamma_{\rho'}(s)} \nonumber\\
-\frac{\delta s e^{i \phi_2}}{M_{\rho''}^2\left[1+\xi_{\rho''} \Re e A_\pi(s)\right]-s-i M_{\rho''} \Gamma_{\rho''}(s)} \,, 
\end{eqnarray}
in which $\xi_\rho=\frac{s}{96\pi^2 F_\pi^2}$ and $\xi_{\rho'}=\frac{s \Gamma_{\rho'}\left(M_{\rho'}^2\right)}{\pi M_{\rho'}^3 \sigma_\pi^3\left(M_{\rho'}^2\right)}$ (analogously for $\xi_{\rho''}$). 
The result for the resummation in Ref.~\cite{hep-ph/0003320} has been employed in the denominator of the $\rho$ contribution.
\section{Other two meson $\tau$ decay channels}
The $\tau^-\to K^-K^0\nu_\tau$ decays are again described only in terms of the vector form factor to an excellent degree of approximation. The current parametrization in 
TAUOLA follows the Guerrero-Pich formula, see Eq.(\ref{GuerreroPich}) \cite{arXiv:hep-ph/9707347, arXiv:0803.2039}. There is also an option to use Eq.(\ref{SU2formula}). Further 
developments in $F^V_{\pi\pi}(s)$ will be immediately translated to $F^V_{KK}(s)$.
The vector form factor in the $\tau^-\to (K\pi)^-\nu_\tau$ decays is currently coded following Ref.~\cite{arXiv:0803.1786}
 \begin{eqnarray} \label{VFFKpi}
F^{V}_{K\pi}(s) &\!\!\! =&\!\!\! \left(\frac{M^2_{K^*}+s\gamma_{K\pi}}{M^2_{K^*}-s-iM_{K^*} \Gamma_{K^*} (s)}
-  \frac{s\gamma_{K\pi}}{M^2_{K^{*\prime}}-s-iM_{K^{*\prime}} \Gamma_{K^{*\prime}} (s)}\right)\times \nonumber\\
&& \mathrm{exp} \left\lbrace\frac{-s}{128\pi^2 F^2} \Re e\left[A_{K\pi}(s) + A_{K\eta}(s)  \right] \right\rbrace . 
\end{eqnarray}
The function $A_{PQ}(s)$ is \cite{ChPT}
\begin{equation}
 A_{PQ}(s) = -\frac{192\pi^2\left[s\,M_{PQ}(s)-L_{PQ}(s)\right]}{s}\,,
\end{equation}
in the notation of Gasser and Leutwyler. An option will be given to switch between this form factor and the one in Ref.\cite{arXiv:0807.4883} \footnote{Other interesting approaches 
are those of Refs.\cite{OtherKpi}.}. The scalar form factor is fundamental in this decay channel to achieve a precise description of the decay data \cite{Kpidata} at low 
values of $s$. Moreover, it is essential to understand CP violation in this channel, which has been reported recently \cite{arXiv:1109.1527}. TAUOLA is ready \cite{CERN-TH-6793-93} 
to handle such $\tau^+$ and $\tau^-$ distinguishing terms. The implementation of $F^S_{K\pi}(s)$ \cite{SFFKpi} is documented in Ref.~\cite{RChLTAUOLAS}.
\section{Three meson $\tau$ decay channels}
The $\tau^-\to(\pi\pi\pi)^-\nu_\tau$ and $\tau^-\to(KK\pi)^-\nu_\tau$ decays have been coded following Refs.~\cite{3mesons}. One- 
and two-resonance exchange diagrams were considered within $R\chi T$ and the appropriate short-distance behaviour was required, yielding sets of compatible relations among 
the Lagrangian couplings in both decays (including $F_VG_V=F_\pi^2$, as in the two meson tau decays). The $\rho'$ resonance was introduced phenomenologically to improve the 
description of the data in the $\tau^-\to(\pi\pi\pi)^-\nu_\tau$ decays. The progress with respect to the earlier description given by the KS model can be appreciated in Fig. 1 
of Ref.~\cite{arXiv:0810.5764}. The inclusion of final state interactions (FSI) in these decays is under study \cite{RChLTAUOLAS}. It should improve the agreement with data in the 
$d\Gamma/ds_{ij}$ distributions, specially at low values of $s_{ij}$.
\section{Conclusions}
A set of form factors based on $R\chi T$ calculations, corresponding to $88\%$ of the hadronic width of the $\tau$ lepton, has been implemented in TAUOLA. 
They are ready for precise confrontation with data gathered at Belle and BaBar (and future Belle II $\&$ Frascati superB facilities). 
In order to obtain the maximum possible information from experiments, the theory input to the MC has to be as accurate as possible with known properties respected ($\chi PT$ 
results at low energies, smooth behaviour of the form factors at short distances, unitarity, analiticity, ...). Still, there are improvements to be done in all modes: 
appropriate inclusion of SU(2) breaking in the $\pi^-\pi^0$ channel, stabilization of $F^S_{K\pi}(s)$, inclusion of excited resonances in the $KK\pi$ modes, and addition of 
FSI (mainly the $\sigma$ effect) in the $3\pi$ mode.\\ \\
I congratulate the organizers of PHI PSI 11 and of the WG on MC Generators for low-energy Physics for their excellent job. I acknowledge funding of the FPA2007-60323 and 
CPAN (CSD2007-00042) grants.






\end{document}